\documentclass[pra,twocolumn,showpacs,preprintnumbers,amsmath,amssymb,superscriptaddress]{revtex4-1}

\usepackage{graphicx}% Include figure files
\usepackage{dcolumn}% Align table columns on decimal point
\usepackage{bm}% bold math
\usepackage[normalem]{ulem} %strikethrough the text with \sout{...}

\usepackage{color}
\usepackage{psfrag}
\usepackage{amsmath}
\usepackage{amssymb}
\newcommand{\abbrev}[1]{\uppercase{#1}}

\newcommand{\kBoltzmann}{k_\text{\abbrev{b}}}

\usepackage{times}

\begin{document}

\title{Phase diagram of Rydberg atoms with repulsive van der Waals interaction}

\author{O.~N.~Osychenko}
\affiliation{Departament de F\'{i}sica i Enginyeria Nuclear, Universitat Polit\`{e}cnica de Catalunya, %
Campus Nord B4-B5, E-08034 Barcelona, Spain}

\author{G.~E.~Astrakharchik}
\affiliation{Departament de F\'{i}sica i Enginyeria Nuclear, Universitat Polit\`{e}cnica de Catalunya, %
Campus Nord B4-B5, E-08034 Barcelona, Spain}

\author{Y.~Lutsyshyn}
\email{yaroslav.lutsyshyn@uni-rostock.de}
\affiliation{Institut f\"ur Physik, Universit\"at Rostock, D-18051 Rostock, Germany}

\author{Yu.~E.~Lozovik}
\affiliation{Institute of Spectroscopy, 142190 Troitsk, Moscow region, Russia}

\author{J.~Boronat}
\affiliation{Departament de F\'{i}sica i Enginyeria Nuclear, Universitat Polit\`{e}cnica de Catalunya, %
Campus Nord B4-B5, E-08034 Barcelona, Spain}

\date{\today}

\begin{abstract}
We report a quantum Monte Carlo calculation of the phase diagram of 
bosons interacting with a repulsive inverse sixth power pair potential,
a model for assemblies of Rydberg atoms in the local van der Waals blockade regime.
The model can be parametrized in terms of just two parameters, the reduced density and temperature.
Solidification happens to the fcc phase.
At zero temperature the transition density is found with the diffusion Monte Carlo method 
at density $\rho = 3.9\;(\hbar^2/m C_6)^{3/4} $, where $C_6$ is the strength of the interaction. 
The solidification curve at non-zero temperature is studied with the path integral Monte Carlo
approach and is compared with transitions in corresponding harmonic and classical crystals.
Relaxation mechanisms are considered in relation to present experiments.
\end{abstract}

%64.70.-p=Specific phase transitions
%67.85.-d=Ultracold gases
%32.80.Ee=Rydberg states 
%64.70.Hz=Phase transitions, solid-vapor

%Use showkeys class option if keyword
%display is desired

\maketitle

\section{Introduction}

Rydberg atoms have one electron excited to a high energy level.
Such atoms exhibit strong and highly tunable interactions which may
have an extraordinarily long range.
Optically excited from suspended clouds of cold atoms,
Rydberg atoms 
interact both between themselves and with the surrounding unexcited atoms, 
resulting in a rich behavior of the Rydberg systems.

Due to the strong interactions, a Rydberg atom shifts
the levels of nearby atoms sufficiently to prevent their subsequent excitation.
A large number of studies
deal with a local blockade regime.
In such a regime a Rydberg atom blocks excitations in its vicinity, 
and the atomic clouds may be injected with well over $10^3$ Rydberg excitations
before the existing excitations block any further ones %
\cite{Phau2007-EvidenceForCoherentCollectiveRydbergExcitationInTheStrongBlockadeRegime,%
Gould2004-LocalBlockadeOfRydbergExcitationInAnUltracoldGas,%
Weidemuller2004-SuppressionOfExcitationAndSpectralBroadeningInducedByInteractionsInAColdGasOfRydbergAtoms}.
Unfortunately, the arrangement of the excited atoms in such experiments is not directly accessible
and has been a subject of intense investigation.
Understanding the ordering of Rydberg atoms may be important 
for interpretation of the experimental results, for example
for the antiblockade effect predicted in \cite{Pohl2007-ManyBodyTheoryOfExcitationDynamicsInAnUltracoldRydbergGas}.
It was also suggested that a spatially
ordered state may allow for a better control over quantum states in such experiments %
\cite{Pohl2010-DynamicalCrystallizationInTheDipoleBlockadeOfUltracoldAtoms}.
Finally, there is an exciting possibility of observing phase transitions in these 
versatile systems, 
especially to states with long-range ordering~%
\cite{Phau2008-QuantumCriticalBehaviorInStronglyInteractingRydbergGases,%
Pfau2009-UniversalScalingInAStronglyInteractingRydbergGas}.

Quantum many-body treatments attempting modeling of realistic Rydberg systems 
have been developed in the past %
\cite{%
Gould2004-LocalBlockadeOfRydbergExcitationInAnUltracoldGas,%
Robicheaux2005-ManyBodyWaveFunctionInADipoleBlockadeConfiguration,%
Pohl2007-ManyBodyTheoryOfExcitationDynamicsInAnUltracoldRydbergGas,%
Cote2010-ManyBodyDynamicsOfRydbergExcitationUsingTheOmegaexpansion,%
Pohl2010-DynamicalCrystallizationInTheDipoleBlockadeOfUltracoldAtoms,%
Pohl2009-MesoscopicRydbergEnsemblesBeyondThePairwiseInteractionApproximation%
}, and were successful in reproducing a number of important experimental features 
\cite{%
Gould2004-LocalBlockadeOfRydbergExcitationInAnUltracoldGas,%
Pohl2009-MesoscopicRydbergEnsemblesBeyondThePairwiseInteractionApproximation,%
Pfau2009-UniversalScalingInAStronglyInteractingRydbergGas,%
Weidemuller2010-EvidenceOfAntiblockadeInAnUltracoldRydbergGas,
Weidemuller2010-CoherentPopulationTrappingWithControlledInterparticleInteractions%
}.
Due to complexity, it is often difficult to consider long-range order with such calculations.
Nonetheless, strong short-range spatial correlations between Rydberg atoms 
were obtained in the calculations of Refs.~\cite{%
Robicheaux2005-ManyBodyWaveFunctionInADipoleBlockadeConfiguration,%
Weidemuller2010-EvidenceOfAntiblockadeInAnUltracoldRydbergGas,%
Cote2010-ManyBodyDynamicsOfRydbergExcitationUsingTheOmegaexpansion%
}, 
as the atoms avoid each other due to the blockade. 
Successful observation of the antiblockade effect
was also a demonstration  of a creation of the strong short-range correlations %
\cite{Weidemuller2010-EvidenceOfAntiblockadeInAnUltracoldRydbergGas}.
Possibility of long-range ordering (crystallization) of Rydberg atoms was recently predicted 
for systems coupled to specially selected chirped laser pulses
\cite{Pohl2010-DynamicalCrystallizationInTheDipoleBlockadeOfUltracoldAtoms,%
Kokkelmans2011-AdiabaticFormationOfRydbergCrystalsWithChirpedLaserPulses}.
Ordering was also considered, and crystalline phase found, in theoretical calculations of both 
one and two-dimensional optical lattices~%
\cite{%
Lesanovsky2010-DynamicalCrystalCreationWithPolarMoleculesOrRydbergAtomsInOpticalLattices,%
Weimer2010-TwoStageMeltingInSystemsOfStronglyInteractingRydbergAtoms,%
Lesanovsky2011-TwoDimensionalRydbergGasesAndTheQuantumHardSquaresModel,%
Markus2011-DislocationMediatedMeltingOfOneDimensionalRydbergCrystals%
}.
Remarkable 
non-commensurate crystalline phases in optical lattices emerged in~Ref.~%
\cite{%
Weimer2010-TwoStageMeltingInSystemsOfStronglyInteractingRydbergAtoms}.

Given the complex nature of the interactions in the Rydberg systems,
it is important to know how much of the behavior of
large assemblies of Rydberg atoms stems directly from the 
pair potential of the interaction between the atoms.
For this reason we aim to study ordering in the simplest model of the Rydberg systems.
Because of the large number of Rydberg-excited atoms in the experiments, we consider the thermodynamic limit. 
While the results are established in the thermodynamic equilibrium,
many present experiments with Rydberg atoms are too short to reach equilibrium. 
Thus comparison in such cases must be made cautiously.

\section{Model and methods}

The dominant interactions in the Rydberg systems are usually the
F\"orster-resonant dipole-dipole interactions between the excited atoms. 
It was shown by Walker and Saffman %
\cite{Saffman2005-ZerosOfRydbergRydbergFosterInteractions, %
Saffman2008-ConsequencesOfZeemanDegeneracyForTheVanDerWaalsBlockadeBetweenRydbergAtoms} %
that, given a pair of Rydberg atoms in the same state, 
the interaction will not have zeros as
a result of the hyperfine structure or alignment of the atoms
only if the resonant coupling is from the $s$ to $p$ states.
Furthermore, interactions in the $s+s\rightarrow p+p$
channels depend only weakly on the hyperfine structure of the $p$ states,
resulting in a nearly isotropic interaction, to within $10^{-2}$.
This perhaps in part motivates the use of the $ns$ Rydberg states in current experiments %
\cite{Phau2007-EvidenceForCoherentCollectiveRydbergExcitationInTheStrongBlockadeRegime%
,Phau2008-RydbergExcitationOfBoseEinsteinCondensates%
,Pfau2009-UniversalScalingInAStronglyInteractingRydbergGas%
,Weidemuller2010-CoherentPopulationTrappingWithControlledInterparticleInteractions}.
Neglecting the hyperfine structure, the interaction for this resonance is isotropic and 
its matrix element is given  in terms of the F\"orster defect $\delta$ as
\cite{Saffman2005-ZerosOfRydbergRydbergFosterInteractions}  
\begin{equation*}
V(r)=\frac{\delta}{2}-\operatorname{sign}(\delta)\sqrt{\left(\frac{C_3}{r^3}\right)^2+\frac{\delta^2}{4}},
\end{equation*} 
which changes from $V = C_3/r^{3}$ 
to van der Waals' $V = C_6/r^{6}$ (with $C_6=-C_3^2/\delta$) for distances much 
larger than the crossover $R_{3\rightarrow 6}\sim(-C_6/\delta)^{1/6}$.
In the case of a strong local blockade, the blockade radius is often larger than the crossover distance.
In such a case, the excited atoms are more likely to be found at 
distances where the interaction is already of the van der Waals type.

The above arguments motivate the repulsive van der Waals model
for the Rydberg atoms in the local blockade regime.
We disregard any energy transfer or interactions with the underlying gas of the ground-state atoms,
and particles are treated as spinless bosons in three-dimensional space with the many-body Hamiltonian
\begin{equation*}
\mathcal{H}=-\frac{\hbar^2}{2m}\sum_i\nabla_i^2+\sum_{i<j}\frac{C_6}{{\left|\bm{r}_i-\bm{r}_j\right|}^6} .
\end{equation*}
Defining the reduced units of length and energy as
\begin{equation}
r_0=\left(\frac{mC_6}{\hbar^2}\right)^\frac{1}{4} %\\
\; , \;
E_0=\frac{\hbar^3}{m^{3/2}C_6^{1/2}}
\label{eq:reducedunits}
\end{equation}
allows us to describe the properties of this model universally in terms of just two parameters,
the dimensionless density $\rho r_0^3$ and temperature $\kBoltzmann T/E_0$.
The units are selected to satisfy $E_0=\hbar^2/mr_0^2=C_6/r_0^6$. 
The mass $m$ in Eq.~(\ref{eq:reducedunits}) is the mass of the atom. 

It is important to establish the applicability of the bulk phase diagram to finite systems. 
For a cloud of size $R$ and
number density $\rho$, the tail potential energy per particle can be estimated as $\rho C_6/R^3$.
In order for the phase transition to occur at the same parameters in the limited system 
as in a bulk one, it is sufficient that the missing potential energy is
much smaller than the kinetic energy.
In the case $T \kBoltzmann\gg E_0$, this reduces to $R\gg \sqrt[3]{\rho C_6 /T \kBoltzmann}$.
When $T \kBoltzmann\ll E_0$, kinetic energy is estimated as $\hbar^2\rho^{2/3}/m$  and thus $R\gg r_0 (\rho r_0^3)^{1/9}$.

\section{Results for the phase diagram of the repulsive van der Waals gas}

\begin{figure}
\includegraphics[angle=270,width=\columnwidth]{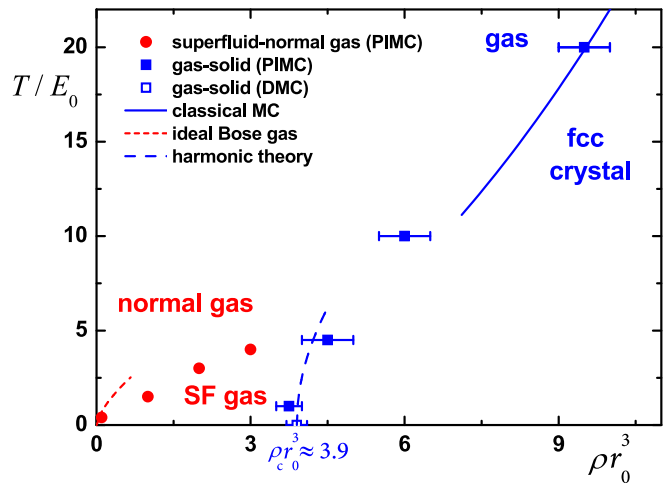}
\caption{\label{fig:phasediagram} (Color online) 
Phase diagram of the repulsive $C_6/r^6$ interaction, scaled to units given in Eq.~(\ref{eq:reducedunits}).
Location of the gas-to-solid transition at zero temperature, determined with \abbrev{dmc}, 
is shown with an open blue square on the $T=0$ axis. 
Dashed blue line shows gas-to-solid transition as 
found with harmonic theory [Eq.~(\ref{eq:Tcharmonic})].
Solid blue line shows classical gas-to-solid transition [Eq.~(\ref{eq:Tclassical})].
Solid blue squares show location of the gas-to-solid transition as determined with \abbrev{pimc}.
Red short-dashed line marks the Bose--Einstein condensation of the ideal gas [Eq.~(\ref{eq:TcBECideal})].
Filled red bullets show Bose--Einstein condensation temperatures found with \abbrev{pimc}.
}
\end{figure}

The phase diagram of the model includes a solid 
at high densities and, at lower densities, a gas phase 
that Bose-condenses at sufficiently low temperatures. 
To locate these phase regions, we employed a number of 
methods, each suitable in a certain area of the phase diagram.
At zero temperature, the model was treated with diffusion Monte Carlo (\abbrev{dmc}), 
a projector method which provides an exact ground-state energy for bosonic systems.
\abbrev{Dmc} has been used successfully in the past to calculate the equations of state and 
locate quantum phase transitions for a variety of systems. 
Transitions at non-zero temperature were studied with path integral Monte Carlo (\abbrev{pimc}), 
a first principles method which allows one to compute the averages 
of quantum operators by summing over the quantum partition function of the system.
Both \abbrev{dmc} and \abbrev{pimc} methods allow one to treat systems with several hundred 
particles under periodic boundary conditions, with thermodynamic limits obtained 
by a suitable extrapolation.
Additionally, classical limits were established with classical Monte Carlo calculations.
In two regimes the location of phase transitions 
could be expressed in a semi-analytical form. 
In the first case, the transition between superfluid and normal gas 
was expressed in terms of the scattering length of the potential by means of a known relationship.
In the second, the solid-to-gas transition was located at low temperatures with the harmonic theory.
The results are summarized in the phase diagram shown in Fig.~\ref{fig:phasediagram}.

At sufficiently high density, the atoms are expected to form a crystalline solid.
Summing the potential energy of the perfect lattice structures, 
we conclude that the preferred symmetry is fcc. 
While other structures may be excluded on the energetic grounds, 
the energy of the hcp structure is very close to that of the fcc.
The difference between the perfect crystal energies,
$E_\text{hcp}-E_\text{fcc}=2 \times 10^{-4} (\rho r_0^3)^2 E_0$,
is small enough to be comparable to or even swamped by the temperature effects 
in present experiments (for example, in Refs.~%
\cite{Gould2004-LocalBlockadeOfRydbergExcitationInAnUltracoldGas%
,Phau2008-RydbergExcitationOfBoseEinsteinCondensates%
,Weidemuller2010-CoherentPopulationTrappingWithControlledInterparticleInteractions}).
The hcp phase is anticipated to be metastable with respect to the transition to the fcc phase.
Zero-point motion and temperature effects are expected to keep the fcc symmetry preferred to hcp.
In the following, the solid phase is always implied to have the fcc structure.

Investigation on the zero-temperature line was done with the \abbrev{dmc} method~\cite{LesterBook1994,Boronat1994}. 
For importance sampling in the gas phase we used a Jastrow form 
$\prod_{i<j} \{\exp\left[ -1/2 (b/r_{ij})^2 \right]+\exp\left[ -1/2 (b/(L-r_{ij}))^2 \right]\}$, 
$r_{ij}<L/2$, for a periodic box of size $L$.
The second power in $1/r$ arises from the cusp condition 
of the scattering problem with the repulsive $1/r^6$ potential
and is also compatible with the presence 
of long-wavelength phonons \cite{Reatto1967-PhononsAndThePropertiesOfABoseSystem}.
The parameter $b$ was variationally optimized beforehand.
The Nosanow--Jastrow wave function was used for importance sampling 
in the solid phase \cite{Nosanow1964,Boronat2008}.
It consists of the product of the above Jastrow term
and a site-localizing Nosanow term $\prod_i \exp\left[-(\bm{r}_i-\bm{l}_i)^2/2\gamma \right]$,
where $\bm{r}_i$ and $\bm{l}_i$ denote correspondingly the coordinates of the
atoms and lattice sites, and $\gamma$ is  
the second optimized parameter. 
The breaking of exchange symmetry between particles in the solid affects the energy only negligibly \cite{Boronat2009NJP}.
Within the statistical errors of the \abbrev{dmc}, results for the energies 
of the fcc and hcp lattices are indistinguishable and both are lower than the energies 
derived using bcc configuration.

While the phase transitions are conventionally reported as a function of pressure rather
than density, density of the Rydberg atoms is more accessible and controllable experimentally.
We therefore choose to express the transition locations in terms of density, 
even for the first-order solidification transition (in this case one needs to specify the coexistence region).
We find that the equations of state for the fcc solid and gas phases cross at the transition density
\begin{equation}
\rho_c \, r_0^3=3.9 \pm 0.2,
\label{eq:TcDMC}
\end{equation}
expressed in the reduced units with the help of Eq.~(\ref{eq:reducedunits}).
The coexistence region of the solid and gas phase at zero temperature, 
determined using the double-tangent Maxwell construction, is narrow 
and is in fact smaller than the above error for the transition density (which arises mostly 
from the extrapolation to the thermodynamic limit; calculations were performed with up to 256 particles).

The transition line between solid and gas phases at small temperatures
can be determined with the harmonic theory \cite{Ashcroft}, 
assuming the Lindemann ratio remains unchanged on the transition line. 
The value of the Lindemann parameter at melting 
may be extracted from the \abbrev{dmc} calculations of the transition density at zero temperature. 
The resulting low-temperature dependence of the gas-to-solid transition density is given by
\begin{equation}
T^{\text{harmonic}}_c=C\sqrt{\left(\rho-\rho_c\right)r_0^3}\frac{E_0}{\kBoltzmann},
\label{eq:Tcharmonic}
\end{equation}
where $\rho_c$ is the transition density at zero temperature, Eq.~(\ref{eq:TcDMC}), and
the constant $C=8.0$ is determined numerically from the dispersion 
curves of the solid and depends on the interactions and geometry of the fcc lattice.

A quantum solid melts at lower temperatures 
than the classical one due to the zero-point motion of the atoms.
The classical transition was located in the canonical ensemble 
by Metropolis sampling of the Boltzmann factor. %
As the potential energy $C_6/r^{6}$ is exactly proportional 
to the square of the density, the transition temperature 
for the classical system also scales exactly as $T\propto\rho^2$. 
We find that 
\begin{equation}
T_c^\text{classical}=0.22\left(\rho r_0^3\right)^2 \frac{E_0}{\kBoltzmann}.
\label{eq:Tclassical}
\end{equation}
As expected, such scaling removes the Planck constant from the classical transition temperature,
which in fact simplifies to $T_c^\text{classical} \kBoltzmann=0.22 \rho^2 C_6$.

To fully account for quantum effects, the gas-to-solid transition at $T\ne 0$
was also located with \abbrev{pimc} calculations. 
We used decomposition of the action operator that is accurate 
beyond the fourth order
\cite{Chin2004-QuantumStatisticalCalculationsAndSymplecticCorrectorAlgorithms}.
For details of the method and implementation, 
see Ref.~\cite{Boronat2009-HighOrderChinActionsInPathIntegralMonteCarlo}. 
The transition was located  by observing 
melting or solidification while working in the canonical ensemble, 
beginning with configurations of atoms placed on a randomly distorted lattice.
Used in this way, the calculations determine a range in which the transition density is located.
\abbrev{Pimc} results confirm the validity of the harmonic approximation at low temperatures.
At higher temperatures the transition density follows the classical melting curve (\ref{eq:Tclassical}).

The above results establish the solidification transition of the repulsive van der Waals model.
Additionally, the dynamic nature of the Rydberg gas
raises a possibility for the spatial ordering to
be induced kinetically, as the combination 
of decay and strong blockade
will favor supplanting excitations to be equidistant
from their immediate neighbors. 
We modelled such a process and observed that 
replacement of decaying excitations in the local blockade regime
indeed creates a short-distance order,
but not a true long-distance crystalline ordering.
These finding are consistent with much more elaborate dynamic models of 
Refs.\ \cite{%
Robicheaux2005-ManyBodyWaveFunctionInADipoleBlockadeConfiguration,%
Weidemuller2010-EvidenceOfAntiblockadeInAnUltracoldRydbergGas,%
Cote2010-ManyBodyDynamicsOfRydbergExcitationUsingTheOmegaexpansion%
}.

At low temperature, the gas phase of the model is expected to form a Bose--Einstein condensate (\abbrev{bec}). 
Transition between the \abbrev{bec} and normal gas phases
at low densities lies slightly above  the ideal Bose gas condensation temperature,
\begin{equation}
T_\text{\abbrev{bec}}^\text{ideal}=2\pi\left(\frac{\rho r_0^3}{2.612\dots}\right)^{2/3}\frac{E_0}{\kBoltzmann},
\label{eq:TcBECideal}
\end{equation} 
due to the repulsive interaction 
between particles \cite{Giorgini2008-CriticalTemperatureOfInteractingBoseGasesInTwoAndThreeDimensions}. 
The correction is governed by the scattering 
length of the potential $a_s$, which can be found to be equal to
$a_s=2\, \Gamma\left(3/4\right)/ \, \Gamma\left(1/4\right) r_0=0.676\dots r_0$.
The transition temperature is then given by 
$
T_\text{\abbrev{bec}}=T_\text{\abbrev{bec}}^\text{ideal}\left(1+c a_s \rho^{1/3} \right)
$,
where $c$ is a positive constant of the order of unity 
(for details, see Ref.~\cite{Giorgini2008-CriticalTemperatureOfInteractingBoseGasesInTwoAndThreeDimensions} and references therein). 
In the present case this expression is only valid at
very low densities 
(one needs to satisfy at least $\rho r_0^3 < 5\times 10^{-2}$ to make the description 
in terms of the zero-momentum scattering length meaningful), 
where the magnitude of the correction is not significant.

At higher densities the \abbrev{bec}-to-normal gas transition is no longer universal 
and depends on the form of the potential. 
We determined the location of this second-order transition with the \abbrev{pimc} method 
by calculating the superfluid transition from the winding number estimator \cite{Ceperley1987-PathIntegralComputationOfSuperfluidDensities}.
The \abbrev{pimc} calculations show that, at higher densities, 
the interactions deplete the condensate 
and the transition temperature is lower than for the ideal Bose gas.
Combining the \abbrev{pimc} results, the region
in which the triple point is located was determined 
as 
$4.5 < T/(E_0/\kBoltzmann)   < 6.5$ and $4 < \rho r_0^3 < 5$, 
which we consider sufficiently narrow for practical considerations.

\section{Comparison with experimental conditions}

Because the interaction constant $C_6$ enters 
the reduced units [Eq.~(\ref{eq:reducedunits})], the effective
temperature and density can be varied over many orders of magnitude.
Most of the present experiments are deeply in the ``classical''
region of the phase diagram (Fig.~\ref{fig:phasediagram}).
As an example, we consider the conditions of the experiments presented 
in Ref.~\cite{Phau2008-RydbergExcitationOfBoseEinsteinCondensates}.
For the excitation with 170~ns laser pulses, the system parameters 
at $4~\mu\text{K}$ are $T/(E_0/\kBoltzmann)\approx33\times10^{5}$ 
and $\rho r_0^3\approx1.9\times10^3$, which in fact correspond to 
the gas phase of the equilibrium phase diagram.
For 320~ns excitation pulses and $T=1~\mu\text{K}$, 
$T/(E_0/\kBoltzmann)\approx8.2\times10^{5}$ and $\rho r_0^3\approx7.4\times10^3$, well below 
the gas-to-solid transition.
Therefore, the achievable temperature and density 
are already in the range suitable for investigating the equilibrium phase diagram.
Increasing the excitation number increases the interaction 
constant $C_6$ and moves the system deeper 
into the classical regime where the gas and solid phases are
separated by the simple condition of Eq.~(\ref{eq:Tclassical}).
The quantum regime of the phase diagram may be accessed by decreasing 
the excitation numbers or increasing the F\"orster defect $\delta$.
 
Whether Rydberg atoms in actual experiments will reach or even approach
an equilibrium phase depends on their lifetime, 
the experiment duration and availability of relaxation mechanisms.
Because of the short lifetimes of the Rydberg states, 
most current experiments are performed on such short 
timescales as to make the thermal motion negligible. 
It is therefore said that the experiments are performed with Rydberg excitations of a frozen gas.
If the experiments are extended closer to the currently achievable 
lifetimes of the Rydberg states, which can be as large as 100~$\mu$s % 
\cite{Robicheaux2008-SpatiallyResolvedObservationOfDipoleDipoleInteractionBetweenRydbergAtoms,%
Pfau2009-UniversalScalingInAStronglyInteractingRydbergGas},
some degree of thermal equilibration will already be achieved.
Besides the thermal motion there are, however, at least two other kinds of motion that may 
need to be considered.
The first one is the motion of the excited atoms due to the strong forces between them. 
The characteristic timescale associated with such a motion is the time that it 
takes for a Rydberg atom to travel the mean distance 
between Rydberg atoms. Given the mean distance $\xi\approx\rho^{-1/3}$  and the imbalance force of the order of $C_6/\xi^6$, this time is given by
\begin{equation}
t_\text{ballistic}\sim\sqrt{\frac{m}{C_6 \rho^{8/3}}}.
\label{eq:ballistictime}
\end{equation}
For a small fixed number of Rydberg excitations time (\ref{eq:ballistictime}) decreases rapidly 
with the excitation number $n$ as $n^{-11/2}$; 
as the local blockade is reached, $\rho \propto C_6^{-1/2}$, and $t_\text{ballistic}$ instead grows as $n^{11/6}$. 
For example, Rydberg systems created by the $1.970~\mu\text{s}$ pulses from $1 \mu K$ gas
in the experiment of Ref.~\cite{Phau2008-RydbergExcitationOfBoseEinsteinCondensates} have 
$t_\text{ballistic}\approx 12~ \mu\text{s}$.
For the setup of Ref.~\cite{Pfau2009-UniversalScalingInAStronglyInteractingRydbergGas},
$t_\text{ballistic}\approx 60~ \mu\text{s}$ while the clouds 
could be successfully studied for as long as $20~ \mu\text{s}$.
Collisional ionization and heating could potentially hamper such relaxation~\cite{Weidemuller2007-ModelingManyParticleMechanicalEffectsOfAnInteractingRydbergGas}.

~

\section{Discussion and conclusions}

As an example, consider the possibility of exploring the supersolidity in Rydberg systems.
Ground state atoms dressed in Rydberg states exhibit weak van der Waals interactions 
at large distances, as described in~\cite{Pohl2010-ThreeDimensionalRotonExcitationsAndSupersolidFormationInRydbergExcitedBoseEinsteinCondensates}.
Supersolidity of such atoms was already investigated 
\cite{Boninsegni2010-SupersolidDropletCrystalInADipoleBlockadedGas%
,Pohl2010-ThreeDimensionalRotonExcitationsAndSupersolidFormationInRydbergExcitedBoseEinsteinCondensates}.
Consider instead a scenario in which the gas is additionally allowed to have a lattice 
of Rydberg excitations.
Such a lattice would in turn impose weak but long-range spatial correlations onto the ground-state atoms.
At the same time, the ground-state atoms may be placed in the \abbrev{bec} regime 
\cite{Phau2008-RydbergExcitationOfBoseEinsteinCondensates}. 
However, it may be impossible to identify which of the atoms was excited within a certain proximity, 
as was demonstrated, for example, by the superatom analysis of the experimental 
results in Refs.~\cite{Phau2007-EvidenceForCoherentCollectiveRydbergExcitationInTheStrongBlockadeRegime,%
Gould2004-LocalBlockadeOfRydbergExcitationInAnUltracoldGas}. 
(However, motion will lead to dephasing of this state 
\cite{Pfau2011-ArtificialAtomsCanDoMoreThanAtomsDeterministicSinglePhotonSubtractionFromArbitraryLightFields}.)
If the atoms are indeed prepared 
in such a mixed state, combining the ground $|g\rangle$ and excited $|e\rangle$ states as $|g\rangle+\alpha|e\rangle$, 
$\bar{N}_g^{-1}\ll\alpha\ll 1$,
then both the lattice-forming and the \abbrev{bec} components are indistinguishable and may be said to be formed by the same atoms.
Therefore, such a system would consist of particles which
would simultaneously break translational symmetry and possess off-diagonal long-range order, 
an epitomic realization of supersolid. 
While our model does not include the light field, the conditions for the phases 
of both excited- and ground-state atoms may be immediately extracted from Fig.~\ref{fig:phasediagram}, 
just with different reduced units for the two species.

In conclusion, it is possible to 
parametrize a model with isotropic van der Waals interactions 
into a universal phase diagram.
We have characterized the phase diagram 
of Rydberg atoms by considering a model of bosons with
repulsive van der Waals interaction, and determined
solidification and Bose--Einstein condensation conditions.
Relaxation mechanisms other than thermal motion should be
considered if one considers Rydberg systems on time scales of several tenth of microseconds.
Finally, it is worth mentioning that 
interactions between Rydberg excitations 
open a possibility of
new supersolid scenarios.

\begin{acknowledgments}

We acknowledge partial financial support by \abbrev{dgi} (Spain) under Grant No.\
\abbrev{fis}2008-04403 and Generalitat de Catalunya under Grant No. 2009-\abbrev{sgr}1003.
G.E.A. acknowledges support from the Spanish \abbrev{mec} through the Ramon y Cajal fellowship program. 

\end{acknowledgments}

\end{document}